\documentclass[pre,floatfix,twocolumn]{revtex4}
\usepackage{amsmath}
\usepackage{amsfonts}
\usepackage{mathrsfs}
\usepackage{textcomp}
\usepackage{graphicx}
\usepackage{epstopdf}
\usepackage{hyperref}
\usepackage{color}
\usepackage{dcolumn}
\usepackage{commath}
\usepackage{url}
\usepackage{gensymb}
\usepackage{siunitx}
\usepackage{doi}

\begin{document}

\title{Lateral inhibition provides a unifying framework for 
spatiotemporal pattern formation in 
media comprising relaxation oscillators}
\author {R. Janaki$^{1,2}$, Shakti N. Menon$^{1}$, Rajeev Singh$^{1,3}$ and Sitabhra Sinha$^{1,4}$}
\affiliation{$^{1}$The Institute of Mathematical Sciences, CIT Campus, Taramani, Chennai 600113, India}
 \affiliation{$^{2}$Department of Theoretical Physics, University of Madras, 
Guindy Campus, Chennai 600025, India}
 \affiliation{$^{3}$Department of Physics, Indian Institute of 
Technology (BHU), Varanasi 221005, India}
\affiliation{$^{4}$Homi Bhabha National Institute, Anushaktinagar, Mumbai 400094, India.}
\begin{abstract}
The collective dynamics seen in a wide variety of chemical, biological and 
ecological systems involve interactions between relaxation oscillators
that typically involve fast activation process coupled with a 
slower inactivation. In this paper, we show that systems of such oscillators 
having distinct kinetics governing local dynamical behavior and whose 
interactions are described by different connection topologies, can exhibit
strikingly similar spatiotemporal patterns when diffusively coupled via
their inactivation component. We explain the apparent universality of
this global behavior by showing that relaxation oscillators interacting 
via lateral inhibition will generally yield two basic classes of
patterns, viz., one comprising one or more clusters of synchronized
oscillators while the other is a time-invariant spatially inhomogeneous 
state resulting from oscillation death. All observed collective states
can be interpreted either as specific instances of these fundamental
patterns or as resulting from their competition. 
\end{abstract}
\maketitle
The ubiquity of patterns in nature has, for several decades, stimulated 
efforts at understanding possible mechanisms that underlie their 
emergence~\cite{Cross1993,Ball1999}. Such patterns can be manifested
in both space and time, with perhaps the most widespread instance of 
the latter being provided by systems that exhibit relaxation 
oscillations~(see, e.g., Refs.~\cite{Buzsaki2004,Lenz2011,PotvinTrottier2016,Guzzo2018}). 
In the simplest scenarios, these can be understood 
as resulting from interactions between an activator component and 
an inhibitory (or inactivation) component that operate at fast and 
slow time-scales respectively~\cite{Izhikevich2000}.
Such oscillators can, in turn, interact with each other, which can
result in non-trivial emergent collective dynamics in systems ranging from
the cell to the food web~\cite{Glass2001,Pikovsky2001}. As competition between neighboring
elements is a recurring motif in such systems, it is natural to 
consider the consequences of lateral inhibition~\cite{Li2015,Collier1996,Sternberg1988,Blakemore1970} in systems of relaxation oscillators.

In this paper we have explored the collective dynamics in a variety of models arising in chemical, biological and ecological 
contexts. The common thread connecting these diverse systems is
that all of them are described by systems of relaxation oscillators
coupled to their neighbors through the diffusion of their inactivation
components. This may, for instance, arise in spatially extended ecological
habitats comprising several patches, with each exhibiting oscillations in
predator and prey populations, where only the predator (acting as the
inactivation component) can move across neighboring patches, e.g., as
in herbivore-vegetation interactions~\cite{Turchin2003}. Experimental realizations of such systems involving oscillating chemical reactions in microfluidic devices have demonstrated the existence of several striking 
patterns~\cite{Toiya2008}. These include anti-phase synchronization
as well as spatially heterogeneous time-invariant patterns. While the latter
resemble stripes generated by the Turing mechanism~\cite{Turing1952}, it has
been analytically demonstrated using a generic model that these result 
from spatially patterned oscillation death~\cite{Singh2013}. Here we show that the patterns seen in these systems can arise in much more general contexts, specificallyinvolving models characterized by different local dynamics and connection topologies, and which describe processes across a wide range of spatial scales. Such ``universality'' arises from the fact that
all patterns generated by such systems can be seen as either specific
manifestations of, or arising through interactions between, two 
basic classes of patterns. Moreover, these fundamental patterns will be observable in any system where lateral inhibition couples oscillators whose dynamics is governed by interactions between components characterized
by widely separate time scales.

\begin{figure}
\includegraphics{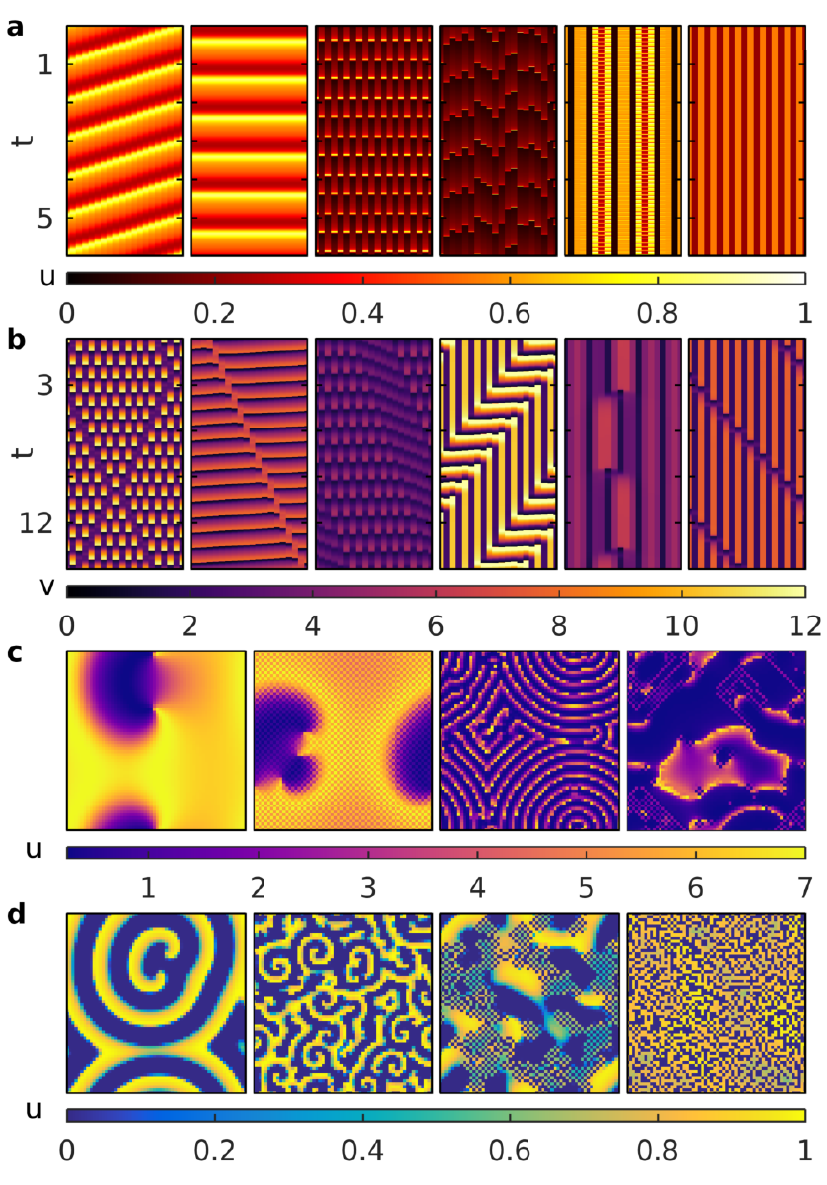}
\caption{(Color online)
Diversity of complex spatiotemporal patterns observed in
several systems described by lattices of
relaxation oscillators, coupled through diffusion of their inactivation
variables.  (a-b) Pseudocolor representation of the spatiotemporal evolution of the activation variable
$u$ for one-dimensional arrays of oscillators described by 
(a) a cell cycle model adapted from Ferrell {\em et al.}~\cite{Ferrell2011}, and (b) the Brusselator model for autocatalytic chemical reactions~\cite{Prigogine1968}.
Both systems comprise $N=20$ oscillators arranged on a ring.  
Attractors of the spatiotemporal dynamics of the system shown in (a) include 
(L-R) Gradient Synchronization (GS), Synchronized Oscillations (SO), a special case of GS, Anti-phase Synchronization (APS),  a pattern exhibiting generalized synchronization, Chimera State (CS) characterized by co-existence of oscillating and non-oscillating elements, and Spatially Patterned Oscillation Death (SPOD) state. Other complex patterns in addition to those mentioned above emerge from the spatiotemporal evolution of the system of coupled chemical oscillators shown in (b). Snapshots of the activation variable $u$ in a two dimensional array (with periodic boundary conditions) comprising $N \times N$ relaxation oscillators described by (c) the Brusselator model, and (d) the Rosenzweig Macarthur model of predator-prey dynamics~\cite{Turchin2003}. For both systems, $N=60$ with each oscillator coupled to their four nearest neighbors. The patterns seen in (c) include (L-R) GS and generalized APS, both exhibiting spiral waves, a tightly wound spiral representing a complex phase relationship between the oscillators and CS (checkerboard regions comprise non-oscillating nodes). 
Similar patterns are seen in (d), viz., (L-R) single and multiple spirals, 
CS and SPOD.
}
\label{fig1}
\end{figure}
\begin{figure}
\includegraphics{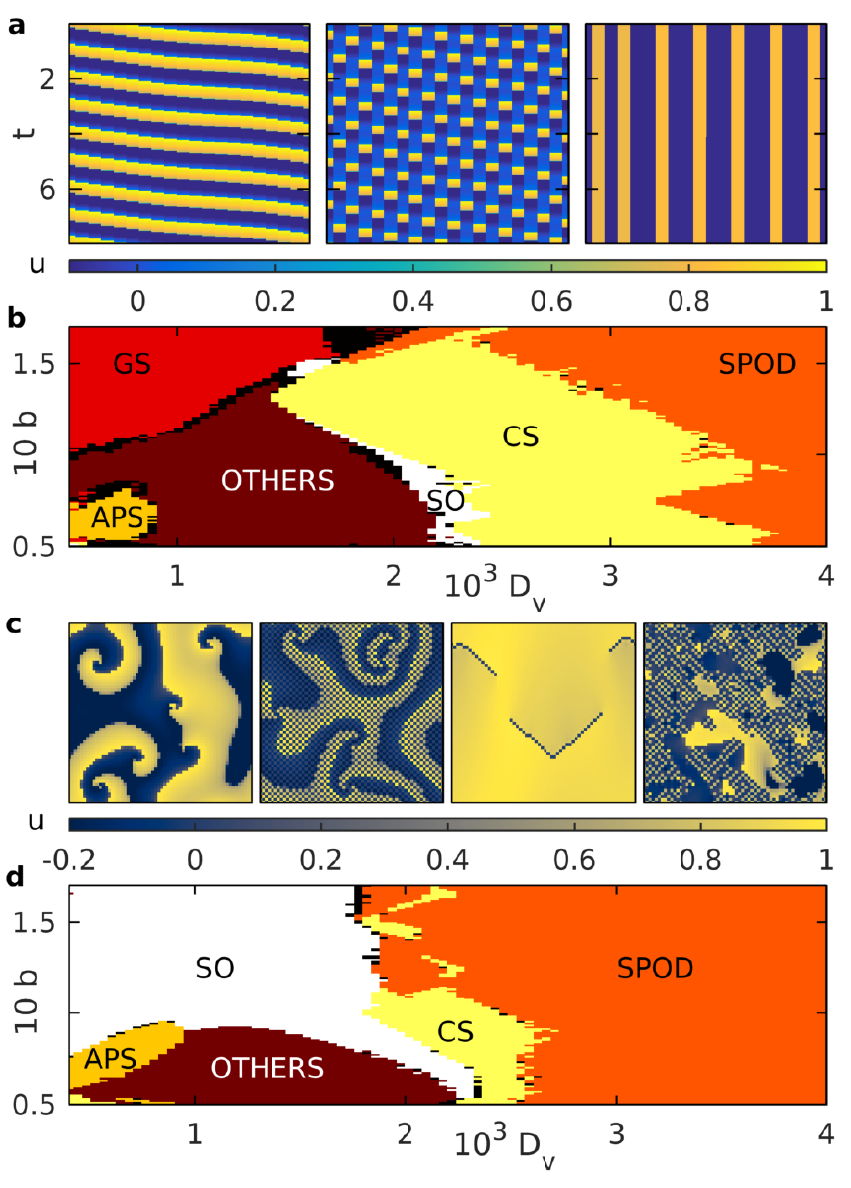}
\caption{(Color online) Collective dynamics in lattices of coupled relaxation oscillators described by the generic Fitzhugh-Nagumo (FHN) model. (a) Pseudocolor representation of the spatiotemporal evolution of the activation variable $u$ for one-dimensional arrays showing
(L-R) Gradient Synchronization (GS, of which Synchronized Oscillations, SO, is a special case), generalized Anti-Phase Synchronization (APS) and Spatially Patterned Oscillator Death (SPOD). Almost all patterns exhibited by the system, as well as those in Fig.~\ref{fig1}(a-b), are either one of these or can be viewed as combinations thereof (e.g., Chimera State, CS). The system comprises  $N=20$ oscillators diffusively coupled on a ring through the inactivation variable $v$ with strength $D_{v}$.  
(b) Different dynamical regimes of the above system in the $D_{v}-b$ 
parameter plane, labelled by the attractor to which the majority 
($>50\%$) of initial conditions converge, viz SO, GS, APS, CS and SPOD, as well as a variety of other patterns that are referred together as OTHERS. If a majority of initial conditions do not converge to a single attractor, the corresponding region is shown in black. (c) Snapshots of the activation variable $u$ in a two dimensional array (with periodic boundary conditions) comprising $N \times N$ relaxation oscillators displaying: (L-R) a GS and a generalized APS state both exhibiting travelling fronts in the form of spiral waves, `gliders' (line-like phase defects propagating on SO background), and CS (checkerboard regions correspond to non-oscillating nodes). (d) Dynamical regimes of the two-dimensional system analogous to the parameter space diagram shown in (b). For (c-d), oscillators are coupled to their four nearest neighbors on the lattice with periodic boundary conditions; for (c) $N=60$ and (d) $N=10$.
}
\label{fig2}
\end{figure}
\begin{figure}[h!]
\includegraphics{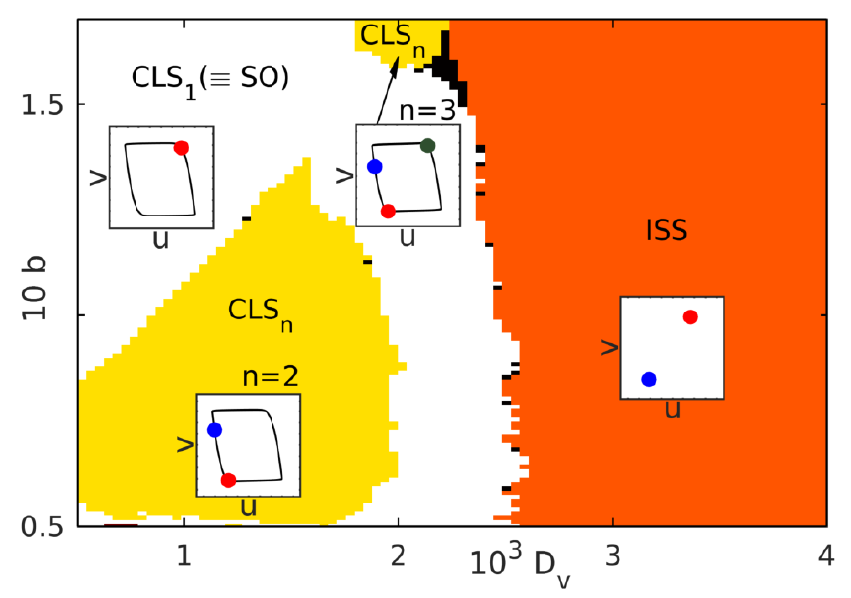}
\caption{(Color online)  
Different dynamical regimes of a globally coupled system of $N (=100)$ FHN
oscillators diffusively coupled through the inactivation variable $v$ are
shown in the $D_{v}-b$ parameter plane. This mean-field model displays
two fundamental patterns of collective activity, viz., Cluster Synchronization (CLS) at low $D_v$ and Inhomogeneous Steady State (ISS) at high $D_v$. The CLS states are further classified on the basis of the number of clusters $n$ into which the oscillators are grouped according to their phase, e.g., CLS$_1$ which is identical to SO, CLS$_2$ which is equivalent to APS and CLS$_3$ that is seen for high $b$.
The insets show the location of each oscillator (colored circles) on
its trajectory in ($u,v$) phase space at a particular time instant for
the dominant attractor in each regime. Note that in ISS the oscillators
are arrested at low or high values, analogous to SPOD seen in lattices. 
}
\label{fig3}
\end{figure}
\begin{figure}
\includegraphics{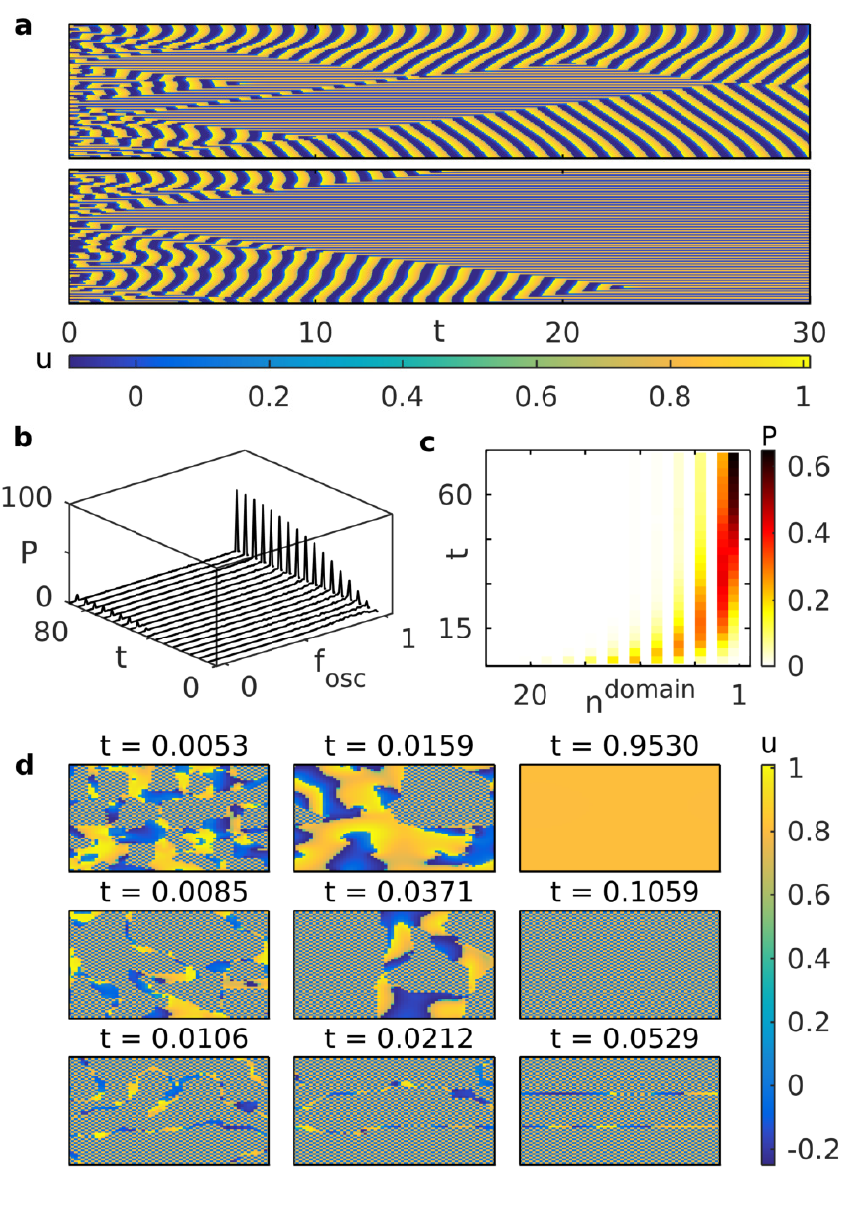}
\caption{(Color online)
Convergence of the collective dynamics of lattices of FHN oscillators 
to one of the fundamental attractors can be understood as a coarsening process. (a) Pseudocolor representation of the spatiotemporal evolution of 
the activation variable $u$ on a ring of $N=100$ oscillators. For the same set of parameter values ($b = 0.175$, $D_v = 1.8 \times 10^{-3}$), the system converges to either GS (top) or SPOD (bottom), depending on the random initial state.
(b) The coarsening shown in (a) can be quantitatively represented 
in terms of the evolution of the Probability density function (pdf P) 
for $f_{osc}$, the fraction of oscillating nodes at a given time. 
Note that asymptotically the distribution converges to an approximately 
bimodal form comprising two peaks around $f_{osc}=0$ (corresponding to SPOD) 
and $f_{osc}=1$ (corresponding to GS). The probabilities are estimated from $10^3$ realizations. (c) The process of coalescence of multiple domains, each comprising either oscillating or non-oscillating nodes exclusively, is 
quantitatively displayed in terms of the time evolution of the pdf for 
the number of such domains, $n^{\rm domain}$. (d) Snapshots of $u$ in a two dimensional array (with periodic boundary conditions) comprising $N \times N$ ($N=60$) relaxation oscillators. For the same set of parameter values as above, the system converges to either SO (top) or SPOD (middle), depending on the initial state. It is also possible to see states where further 
coarsening to SPOD is arrested by the presence of line defects, comprising oscillating cells, that move extremely slowly (bottom).
}
\label{fig4}
\end{figure}
The dynamical behavior of a spatially extended system of
relaxation oscillators that interact over a general connection topology
can be described by the following system of ordinary differential equations:
\begin{equation}
\begin{split}
d u_{i}/dt &= \mathcal{F}(u_{i}, v_{i}) + D_{u} \Sigma_{j \in S_{i}} (u_{j}-u_{i})/k_i\,,\\
d v_i / dt &= \mathcal{G}(u_{i}, v_{i}) + D_{v} \Sigma_{j \in S_{i}}(v_{j}-v_{i})/k_i\,,
\end{split}
\label{eq:RDframework}
\end{equation}
where $u_{i}$ and $v_{i}$ ($i = 1, 2, \ldots N$) represent the activation 
and inactivation components, respectively. 
The dynamics of an
uncoupled node is specified by the functions $\mathcal{F}$
and $\mathcal{G}$, and are governed by parameters whose values are chosen
so as to yield oscillations.
The diffusion terms 
represent interactions of each node $i$ with its neighbors
(comprising the set $S_i$ having size $k_i$) through $u$ and $v$ with
coupling strengths $D_{u}$ and $D_{v}$ respectively. The net 
contribution that each node receives through diffusive interactions
is normalized by the number of its neighbors ($k_i$) so as to make the results
comparable across systems with different connection topologies.
In this paper we focus on spatiotemporal patterns seen in systems 
possessing simple regular geometries, viz., one and two-dimensional lattices.
Motivated by examples of lateral inhibition in nature, where communication between neighboring regions occurs almost exclusively through the inactivation component (such 
as the ones mentioned earlier), we consider the case $D_{u}=0$ for 
all simulations reported here (unless mentioned otherwise).
We have explicitly verified that qualitatively similar results are
obtained even when this constraint is relaxed (e.g., when $D_u = D_v$) 
provided the kinetics of the inactivation component is
sufficiently slower than that of the activation component~\cite{SI}.
For most results reported here, $N = 20$ (for 1-D) and $60^2$ (for 2-D), 
although we have used other values of $N$ to verify that our results are not
sensitively dependent on system size. The boundary conditions are taken
to be periodic in order to minimize boundary effects.
The equations are solved using a variable step stiff solver.
Time units in each case are normalized with respect to the time period of an
uncoupled oscillator for the corresponding set of parameter values. 

We have carried out simulations with different dynamical systems
describing the behavior of individual nodes, corresponding
to cell-cycle oscillations (adapted from Ferrell {\em et al.}~\cite{Ferrell2011}), 
chemical kinetics (viz., the Brusselator model~\cite{Prigogine1968}) and 
predator-prey interactions (belonging to the Rosenzweig-MacArthur class
of models~\cite{Rosenzweig1963}, see Ref.~\cite{Turchin2003}).
Despite very different expressions for $\mathcal{F}$ and $\mathcal{G}$ 
representing the intrinsic activity for each of these systems, 
upon coupling the oscillators exhibit strikingly similar patterns (Fig.~\ref{fig1}).
Furthermore, the broad nature of the collective dynamics does not appear to
depend appreciably on the dimensionality of the lattice, or indeed,
its connectivity.
The diversity of patterns that can be generated by these models is
illustrated in Fig.~\ref{fig1}(a) and (b), corresponding to spatiotemporal
activity in a ring of cell-cycle oscillators and a chemical system
undergoing auto-catalytic reactions that exhibits periodic activity
far from equilibrium, respectively.
The different patterns shown in (a) correspond to different arrangements 
of phase relations between neighboring oscillators that are seen
for a broad range of system parameters. These include (i) Gradient 
Synchronization (GS), characterized by a monotonic change in the phase of 
oscillators along the array (manifested as a propagating front of activity), 
a special case of which is (ii) Synchronized
Oscillations (SO) in which all oscillators are in the same phase, (iii) 
Anti-Phase Synchronization (APS) where the phase of neighboring oscillators 
differ by $\pi$, 
(iv) states corresponding to generalized phase
synchronization, (v) Chimera State (CS) comprising co-occurring
oscillating and non-oscillating nodes, and (vi) Spatially Patterned 
Oscillator Death (SPOD) in which the oscillators are arrested at
different phases. 
Rings comprising oscillators described by the Brusselator model
shows patterns similar to those seen in (a) as well as
more complex ones~\cite{SI}. 
As shown in (b), these include one or more {\em phase defects}, viz., 
a local phase 
relation between a set of
neighboring oscillators that is distinct from the rest of the lattice,
moving across the domain.
These relatively exotic patterns occur over a restricted range of the 
parameter space and are specific to the nature of $\mathcal{F}$ and
$\mathcal{G}$.

Even for higher spatial dimensions, we observe that the
collective dynamics of systems characterized by different
relaxation oscillator models exhibit similar characteristics. 
Specifically, on comparing the
spatiotemporal activity of two-dimensional lattices of coupled
Brusselators [Fig.~\ref{fig1}(c)] and interacting predator-prey populations
described by the Rosenzweig-MacArthur model [Fig.~\ref{fig1}(d)],
a common set of patterns is observed. These include higher dimensional
analogues of GS shown above for 1-D lattices, as well as, generalized
APS (where the phase of an oscillator at time $t$ differs by $\pi$ from
that of its neighbor at $t+\delta t$, where $\delta t$ is a short time delay).
In both of these cases, we observe that 
the propagating fronts take the form of spiral waves.
For stronger diffusive couplings in both systems, we observe CS and SPOD 
patterns analogous to those reported above for 1-D systems.
This suggests that the effective lateral inhibition implemented by
diffusion through the inactivation variable is the dominant
factor underpinning the patterns that can be seen in these very different
systems. 

In order to investigate in detail the collective dynamical phenomena
common to the systems of relaxation oscillators shown in Fig.~\ref{fig1}
we consider a generic model of such oscillators, viz., the
Fitzhugh-Nagumo (FHN) model, to describe the dynamics of each node.
As in Eqn.~(\ref{eq:RDframework}), each oscillator is described by a 
fast activation variable $u$ and a slow inactivation variable $v$. Their 
time-evolution is governed by the functions
$\mathcal{F}(u, v) = u (1 - u) (u - \alpha) - v$ and $\mathcal{G}(u, v)
= \epsilon (k u - v - b)$,
where $\alpha=0.139$, $k = 0.6$ are parameters specifying the kinetics, 
$b$ characterizes the asymmetry (related to the ratio of the time the 
oscillator spends in the high- and low-value branches of the $u$ nullcline)
and $\epsilon = 10^{-3}$ is the recovery rate.
We have verified that the results reported here are not sensitive
to small variations in these values. Moreover, introducing different
boundary conditions can yield qualitatively similar results~\cite{Singh2013}.

Fig.~\ref{fig2}~(a) shows the three most general patterns that can be
observed in a ring of FHN oscillators, viz., GS, generalized APS and SPOD. 
In addition to these, the system exhibits all other robust patterns that
are seen over a wide range of parameter values in the systems described 
earlier [see Fig.~\ref{fig1}~(a-b)] which can be viewed as 
either special cases (such as SO and APS) or combinations of the three 
aforementioned general patterns. For instance, CS corresponds to
part of the system being in GS while the remainder converges to SPOD.

To quantitatively analyze the robustness of the observed patterns over the
($b,D_v$) parameter space, we have estimated the size of their respective
basins of attractions from many ($\sim 10^3$) realizations 
[Fig.~\ref{fig2}~(b)]. In order to classify the space into distinct
pattern regimes we have used the following order parameters. 
First, SPOD and CS states are distinguished by determining the number
of nodes for which the temporal 
variance of the activation variable, $\sigma^{2}_{t}(u_{i})$ is zero. This 
allows us to define the number of non-oscillating cells $N_{no}$ that
is used to distinguish between SPOD ($N_{no} = N$) and CS ($0<N_{no}<N$).
To distinguish between states where all nodes 
are oscillating, including SO, APS and GS, we obtain the time-average of the variance of the activation variable calculated over the lattice, $\langle \sigma^{2}_{i}(u)\rangle_{t}$, as well as the corresponding quantity for each of the two sublattices comprising alternating sites. The latter are zero for both SO and APS states, which are then distinguished by determining whether $\langle \sigma^{2}_{i}(u)\rangle_{t}$ is zero (SO) or not (APS).
If all three time-averaged variances have finite values, the state is
characterized as GS if the times at which the activation
variable of different nodes reaches the peak value changes monotonically
over the lattice, else it is classified as OTHERS.
Note that OTHERS includes a large number of diverse complex spatiotemporal 
patterns including generalized APS. In practice, the different pattern regimes are identified by specifying thresholds on the above order parameters, whose specific values do not affect the qualitative nature of
the results. Parameter regions are marked as GS, SO, APS, SPOD, and CS states if they are obtained for $>50\%$ of random initial conditions (i.e., have the largest basin). While changing the system size do not broadly affect the qualitative features, we note that certain patterns such as SO are harder to obtain in larger systems (as the rapid coordination required for exact synchronization is easier to achieve for smaller systems), while the basins of other patterns (such as APS) shrink considerably. 

A striking feature of the ($b,D_v$) parameter space diagram for the
1-dimensional array of coupled FHN oscillators is that the three
general patterns occur at either end of the coupling strength range, with
SPOD being found at the higher values of $D_v$ while in the lower range
we observe GS as well as anti-phase patterns. The anti-phase collective
dynamics manifests itself when the individual oscillator limit cycles 
are highly asymmetric, corresponding to lower values of $b$. 
Indeed, in the limit of extreme asymmetry where a node remains in one
of the branches (slow or fast) for almost the entire duration of its
oscillation period, it can be shown analytically that APS is the only 
stable state for the system~\cite{Singh2013}.
We would also like to note that while SPOD states resemble Turing 
patterns~\cite{Turing1952}, the generative mechanism is quite distinct from that of Turing instability and involves the arrest of oscillators into
a heterogeneous stationary state, as demonstrated in Ref.~\cite{Singh2013}.
Consistent with the earlier statement that all patterns other than the
three general ones (and their special cases) can be seen as combinations
thereof, we observe that these patterns (such as CS) occur in the 
region between the GS and SPOD regimes. Observation of SO at higher values of $D_v$ can be interpreted as a result of increased coordination resulting from stronger coupling between the oscillators.

Investigation of the spatiotemporal dynamics in two-dimensional lattices 
of $N \times N$ coupled FHN oscillators reveals the existence of patterns 
similar to those seen in Fig.~\ref{fig1}~(c-d), including spiral waves, propagating phase defects and CS [Fig.~\ref{fig2}~(c)].
The corresponding $(b,D_v)$ parameter space [Fig.~\ref{fig2}~(d)] displays
regimes corresponding to SO, APS, CS, SPOD and OTHERS identified using
methods similar to those used for 1-dimensional lattices.
The strong qualitative similarity with Fig.~\ref{fig2}~(b) is visually
apparent. Note that SO is seen to occur over a large region of the parameter space to the exclusion of GS as Fig.~\ref{fig2}~(d) is for a small lattice (viz., $N=10$). For larger lattices, signals take longer to traverse the domain making global phase coherence less likely, which results in localized 
phase coordination manifesting as waves (i.e., GS will dominate).

The qualitative similarity of the parameter space diagrams for $1-$ and 
$2-$dimensional lattices [Fig.~\ref{fig2}~(b) and (d)] 
suggests that the nature of the
pattern regimes seen for a system of such coupled oscillators is 
independent of the dimensionality. To verify this we now consider a
mean-field system of globally coupled FHN oscillators which 
corresponds effectively to the limit of extremely large number of dimensions (Fig.~\ref{fig3}). We observe that the parameter space is dominated by essentially two collective dynamical states, viz., Cluster Synchronization (CLS$_n$) comprising in general $n$ oscillator clusters (each cluster being
characterized by the common phase of all the constituent nodes) and
the temporally invariant Inhomogeneous Steady State (ISS) where the
dynamics of each node is arrested to one of two possible values
[see inset in Fig.~\ref{fig3}]. We would like to point out that all the
observed spatiotemporal patterns mentioned earlier can be viewed
as instances or combinations of these two fundamental states.
In particular, ISS is equivalent to the SPOD state observed 
in a finite-dimensional lattice. The spatially symmetric SO state
where all oscillators have the same phase, and hence belong to
a single cluster, corresponds to CLS$_1$. Similarly, the 
spontaneously broken spatial symmetry APS state comprising two clusters of
oscillators that are exactly $\pi$ out of phase, belongs to CLS$_2$.
We also observe other CLS$_n$ states corresponding to higher values of $n$ 
in small regions of the parameter space. 

Deviating from the mean-field situation by reducing the number of connections
per node will result in the emergence of other robust patterns. 
In particular, it is possible to observe collective states where 
the number of clusters is equal to the total number of nodes
in the system, i.e., CLS$_N$. This will correspond to GS in lattices
with finite coordination number. Its occurrence is inversely related to the
communication efficiency, i.e., how rapidly signals 
coordinate activity across the system. This is governed by the
diffusion strength, as well as, the number of connections (relative
to the system size), and increasing either may result in merging
of clusters that could possibly lead to the globally coherent SO
(i.e., CLS$_1$) state.

As it is now apparent that the observed patterns in the $1-$ and 
$2-$dimensional lattices can be understood as instances of the fundamental
patterns CLS$_n$, ISS and combinations thereof, we now focus on 
how these two collective dynamical states
compete with each other at the interface of these two regimes in
the FHN parameter space. Fig.~\ref{fig4} shows that close to this boundary the system can converge to either one of the two attractors depending on the (randomly chosen) initial condition. This convergence happens after a period
of transient activity that resembles diffusion-mediated coarsening
phenomena seen, e.g., in binary mixtures~\cite{Wen1996,Mullin2000}. As seen in Fig.~\ref{fig4}~(a), a ring of FHN oscillators in this parameter regime
exhibit the rapid creation of several domains of varying sizes, each being
either in a CLS or SPOD state. Over time some of these domains expand at
the expense of others in a process of competitive growth akin to
Ostwald ripening~\cite{Raatke2002}, with the entire system eventually
converging to either CLS$_n$ or SPOD states [top and bottom
panels of Fig.~\ref{fig4}~(a), respectively]. The sizes of the basins of 
attraction for these two states can be discerned from the asymptotic 
probability density of $f_{osc}$, viz., the fraction of nodes that 
belong to any of the domains exhibiting oscillations, whose
time evolution is shown in Fig.~\ref{fig4}~(b).  
Fig.~\ref{fig4}~(c) shows how, as the system approaches the asymptotic state, the number of distinct domains $n^{\rm domain}$ reduces over time
through a process of coalescence. Similar coarsening phenomena leading to any of the two fundamental patterns are also seen for two-dimensional lattices of coupled oscillators [Fig.~\ref{fig4}~(d)].

To conclude, we have shown that a variety of similar patterns are 
exhibited by diverse systems having different local dynamics and connection
topologies. This can be explained by noting that all of these patterns 
are either particular manifestations, or arise through interactions between, two fundamental classes of collective dynamical states. These are characterized either by one or more synchronized clusters, or temporally invariant inhomogeneous patterns. As we show, in general, they will arise from the collective dynamics of a system comprising relaxation oscillators that are coupled through the inactivation components, a feature that is common across the systems that we have considered here. While weak interactions typically generate CLS$_n$, stronger coupling yields ISS. This mechanism of pattern formation, distinct from the classic Turing paradigm, is sufficiently generic to have been observed in experimentally realizable settings, such as in coupled electronic circuits~\cite{Gambuzza2014}. This opens up the possibility of exploring applications for the dynamics reported here, e.g., in the context of computation~\cite{Menon2014}. Finally, as diffusion is not the only mechanism through which the dynamical components of a spatially extended system interact, it will be of interest to see how the introduction of other processes, such as advection, will affect the nature of the spatiotemporal patterns generated by such systems. This will, for example, be relevant in the context of predator-prey systems embedded in environments subject to hydrodynamic flows~\cite{Neufeld2001} (e.g., oceanic plankton populations~\cite{Abraham1998,Neufeld2002}).

\begin{acknowledgments}
RJ is supported by IMSc Project of Interdisciplinary
Science \& Modeling (PRISM) and SNM is supported by 
IMSc Complex Systems Project (XII Plan), both funded by 
the Department of Atomic Energy, Government of India. 
RJ would like to thank A S Vytheeswaran and the Department
of Theoretical Physics, University of Madras for their
support. The simulations required for this work were done in the IMSc
High Performance Computing facility (Nandadevi, Annapurna and Satpura
clusters). The Satpura cluster is partly funded by the Department of Science and Technology, Government of India (Grant No. SR/NM/NS-44/2009) 
and the Nandadevi cluster is partly funded by
the IT Research Academy (ITRA) under the Ministry of
Electronics and Information Technology (MeitY), Government
of India (ITRA-Mobile Grant No. ITRA/15(60)/DIT NMD/01). 
We thank Bulbul Chakraborty, Pranay Goel, Chittaranjan Hens, 
Sandhya Koushika, Sandeep Krishna, Tanmay Mitra, Prasad Perlekar
and Soling Zimik for helpful discussions.
\end{acknowledgments}

\clearpage
\onecolumngrid
\begin{center}
{\large {\bf SUPPLEMENTARY MATERIAL}}
\end{center}
\setcounter{figure}{0}
\setcounter{equation}{0}
\renewcommand\thefigure{S\arabic{figure}}
\renewcommand\thetable{S\arabic{table}}
\renewcommand{\thesection}{\Roman{section}} 
\renewcommand{\thesubsection}{\thesection.\Alph{subsection}}

\section{Model description}
In the main text we have described the collective dynamics of a system
of coupled relaxation oscillators, which are interacting with each
other over a 
specified connection topology (viz., a ring, a two-dimensional torus or 
a globally coupled scenario). The spatiotemporal activity of such
a system of $N$ oscillators, each of which involves $R$ variables, 
can be described by a model involving $N \times R$ coupled differential
equations, viz.,
\begin{equation}
\frac{d x^{(p)}_{i}} {d t} = F_{(p)} (x^{(1)}_i,x^{(2)}_i, \ldots , x^{(R)}_i) 
+ D_{(p)}\sum_{j \in S_{i}}\frac{x^{(p)}_j-x^{(p)}_i}{k_i}\,,\\
\label{eq:1}
\end{equation}
where $x^{(p)}_{i}$ ($i=1,2, \ldots, N$, $p=1,2, \ldots, R$) represents the 
$p$-th component of the $i$-th oscillator. The dynamics of 
an uncoupled node is 
specified by the functions $F_{(p)}$
and governed by parameters whose values are chosen 
so as to yield oscillations. The diffusion terms represent the 
interactions of each node $i$ with its neighbors (which comprises the 
set $S_i$ of size $k_i$) through the different components $x^{(p)}$
with coupling strengths $D_{(p)}$. The net contribution 
that each node receives through these diffusive interactions is 
normalized by the number of its neighbors ($k_i$) so as to make the 
results comparable across systems with different connection topologies.

With the exception of one of the models (viz., the cell-cycle model, described 
below in I.A), the individual oscillators that we have considered are described
by only two components that are responsible for 
activation and inhibition, respectively. These 
systems can hence be described by the equations:
\begin{equation}
\begin{split}
\frac{d u_{i}} {d t} &= \mathcal{F}(u_{i}, v_{i}) + D_{u}\sum_{j \in S_{i}}\frac{u_j-u_i}{k_i}\,,\\
\frac{d v_{i}} {d t} &= \mathcal{G}(u_{i}, v_{i}) + D_{v}\sum_{j \in S_{i}}\frac{v_j-v_i}{k_i}\,,
\end{split}
\label{eq:2}
\end{equation}
which corresponds to Eq.~(1) of the main text. 
In the following sub-sections we describe in detail each of the 
four models whose collective dynamics is reported in the main text.

\subsection{Cell-Cycle model}

In the first system that we consider, the individual oscillators
describe the cell cycle, the periodic sequence of events in a cell
resulting in it dividing into two daughter cells, 
during early embryogenesis in {\em Xenopus laevis}, a
frog that is native to sub-Saharan Africa.
We have adapted the oscillator description used here from a 
three-component model 
developed by Ferrell {\em et al.}~\cite{Ferrell2011a}
which involves interactions between the 
proteins CDK1, Plk1 and APC. In the course of the cell cycle, CDK1 
activates Plk1, which in turn 
activates the protein APC that subsequently suppresses CDK1. 

Representing 
the concentrations of CDK1, Plk1 and APC by $u$, $v$ and $w$, 
respectively, the time-evolution of this system is governed by Eqn.~\ref{eq:1} 
with the functions given by:
\begin{equation}
\begin{split}
F_{u}(u, v, w) &= \alpha_{1}\, -\beta_{1} u\left(\frac{w^{n_{1}}}{k^{n_{1}}_{1} +w^{n_{1}}}\right)
+ \alpha_{4}\, (1-u)\left(\frac{u^{n_{4}}}{k^{n_{4}}_{4} +u^{n_{4}}}\right)\,, \\
F_{v}(u, v, w) &= \alpha_{2}\ (1-v) \left(\frac{u^{n_{2}}}{k^{n_{2}}_{2} +u^{n_{2}}}\right) - \beta_{2}\, v\,, \\
F_{w}(u, v, w) &= \alpha_{3}\ (1-w) \left(\frac{v^{n_{3}}}{k^{n_{3}}_{3} +v^{n_{3}}}\right) - \beta_{3}\, w\,. 
\end{split}
\end{equation}
Note that this system, 
with positive and negative feedback loops, behaves like a 
relaxation oscillator with distinct fast and slow phases. 
Typical values of the model parameters which generate oscillations are
$\alpha_{1}= 0.02$, $\alpha_{2}= 3$, $\alpha_{3}= 3$, $\alpha_{4}= 
3$, $\beta_{1}= 3$, $\beta_{2}= 1$, $\beta_{3}= 1$, $k_{1}= 0.5$, 
$k_{2}= 0.5$, $k_{3}= 0.5$, $k_{4}= 0.5$, $n_{1}=8$, $n_{2}=8$, $n_{3}=8$ 
and $n_{4}=8$.

As APC is a relatively large regulatory complex, one may consider
interactions through diffusion of this protein to be negligible.
We use the fact that Plk1 indirectly suppresses CDK1 by activating APC, to
consider it as the inactivation variable in our simulations. 
We have therefore chosen the diffusion strengths $D_u$ and $D_w$ to be $0$ 
and considered only diffusive coupling through the variable $v$. 
The patterns shown in ~Fig.1(a) of the main text are obtained by varying 
the model parameters $\alpha_1, \alpha_2, \beta_1$ and the
coupling strength $D_v$. 

\subsection{Brusselator chemical oscillator model}
\begin{figure}[h!]
\begin{center}
\includegraphics[width=0.95\columnwidth]{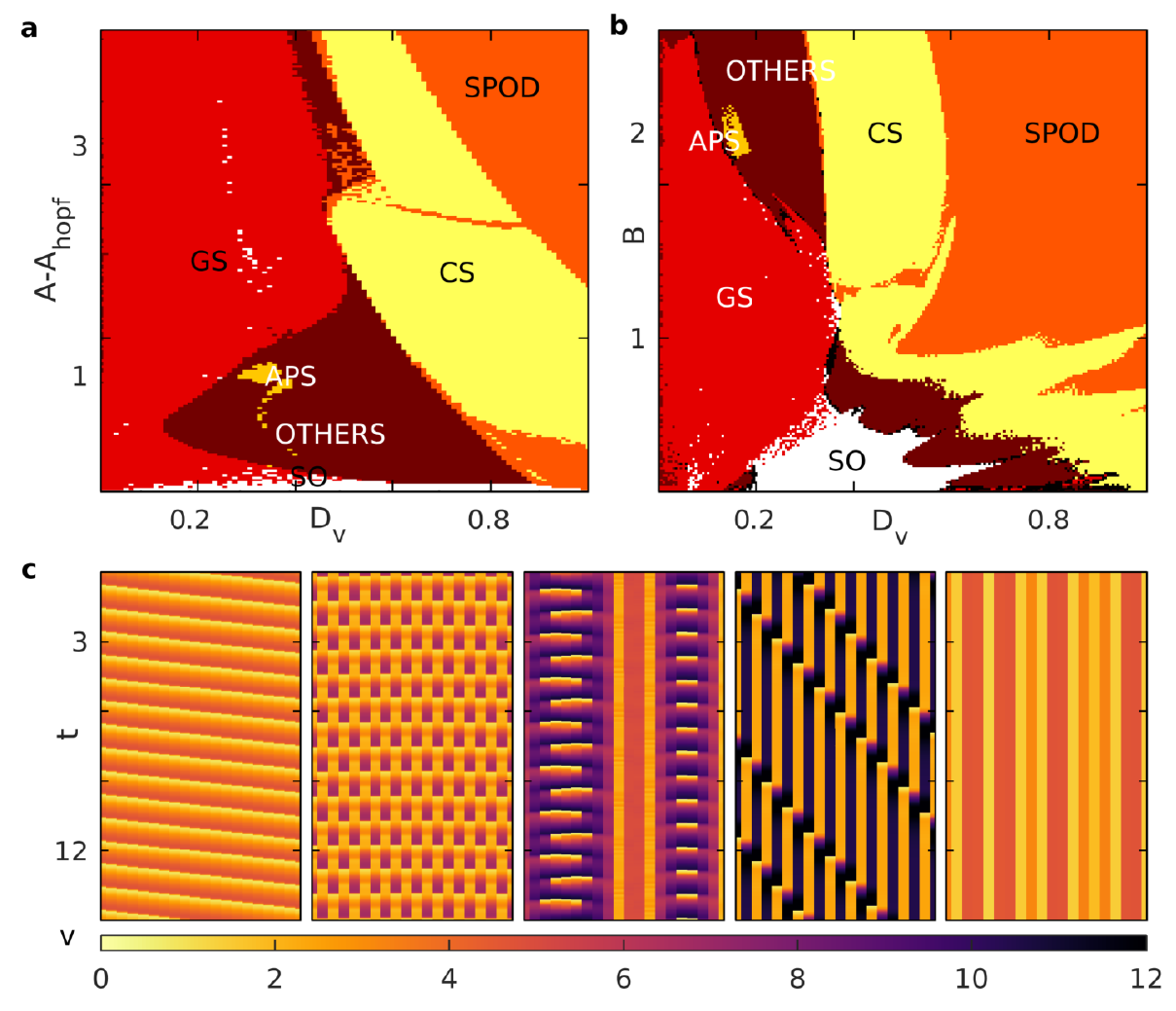}
\end{center}
\caption{
Complex spatiotemporal patterns arising in one-dimensional arrays of 
oscillators described by the Brusselator model, coupled through the 
inactivation parameter $v$ (i.e., $D_u = 0$). 
Different dynamical regimes in (a) 
$D_{v} - (A - A_{hopf})$ parameter plane, and (b) $D_{v} - B$ 
parameter plane are shown, labelled by the attractor to which the majority $(> 
50\%)$ of initial conditions converge, viz. SO, GS, APS, CS, SPOD and 
OTHERS (see main text). (c) Pseudocolor representation of the spatiotemporal evolution 
of the inactivation variable $v$ for one-dimensional arrays of ($N = 
20$) oscillators described by the Brusselator model arranged on a 
ring. In addition to the spatiotemporal patterns displayed in 
Fig.~1(b) of the main text, the system can exhibit other patterns 
including (L-R) GS, APS, CS, travelling SPOD and SPOD.
}
\label{SI1}
\end{figure}

The next system we consider consists of oscillators that describe the 
far-from-equilibrium behavior of chemical systems in which the concentrations
of some reactants exhibit periodic variations.
The specific model used is the {\em Brusselator}~\cite{Prigogine1968a},
a simplified description of autocatalytic chemical reactions such
as the Belousov-Zhabotinsky reaction. 
The time evolution of the system is described by Eqn.~\ref{eq:2} with the
functions $\mathcal{F}$ and $\mathcal{G}$ having the following form:
\begin{equation}
\begin{split}
\mathcal{F}(u, v) &= B + u^{2}\ v - (1+A)\ u, \\
\mathcal{G}(u, v) &= A\ u - u^{2}\ v\,,
\end{split}
\end{equation}
where $A, B$ are parameters whose values can be appropriately chosen
in order to make the system oscillate. 
For our simulations, we consider $A>A_{\rm hopf}(=1+B^2)$, i.e., the
parameter regime in which individual elements have undergone a Hopf bifurcation
to oscillatory behavior.
As mentioned in the main text, in experiments done with Belousov-Zhabotinsky 
reagents in beads that are separate by columns of oil in microfluidic devices,
it is known that only the inactivation variable can effectively 
diffuse through the oil.
We have therefore considered $D_u = 0$. Fig.~\ref{SI1} shows the different
patterns that are obtained by varying the values of the
parameters $A$, $B$ and $D_v$.

\subsection{Rosenzweig-MacArthur predator-prey model}
We next consider a model system whose individual oscillators describe
the dynamics arising from interactions between a prey (whose
population defines the fast activation variable $u$) and a
predator species 
(whose population is given by the slow inactivation variable $v$). 
The specific form for the functions are obtained from the
Rosenzweig-MacArthur model~\cite{Rosenzweig1963a,Turchin2003a}, viz.,
\begin{equation}
\begin{split}
\mathcal{F}(u, v) &= r\, u\left(1-\frac{u}{K}\right) - q\, \frac{u}{b+u}\ v\,, \\
\mathcal{G}(u, v) &= \epsilon\ q\ \frac{u}{b+u}\ v - d\, v\,,
\end{split}
\end{equation}
where $r$ is the intrinsic growth rate, $K$ is the carrying capacity 
of the prey population, $q$ is the maximum predation rate of the 
predator, $b$ is the half saturation constant and $\epsilon$ and $d$ 
represent the efficiency and the death rate of the predator, 
respectively. For our simulations, we have set $r = 1$, $K 
= 1$, $q = 2$, $d=0.1$ and $\epsilon=0.1$. There are many trophic
interactions in which the prey is immobile (e.g., plants) and the
predator is able to graze by moving from one patch to another (e.g., herbivore).
In such a context, one can set $D_u = 0$ and vary $D_v$.

\subsection{FitzHugh-Nagumo model}
\begin{figure}[h!]
\begin{center}
\includegraphics[width=0.95\columnwidth]{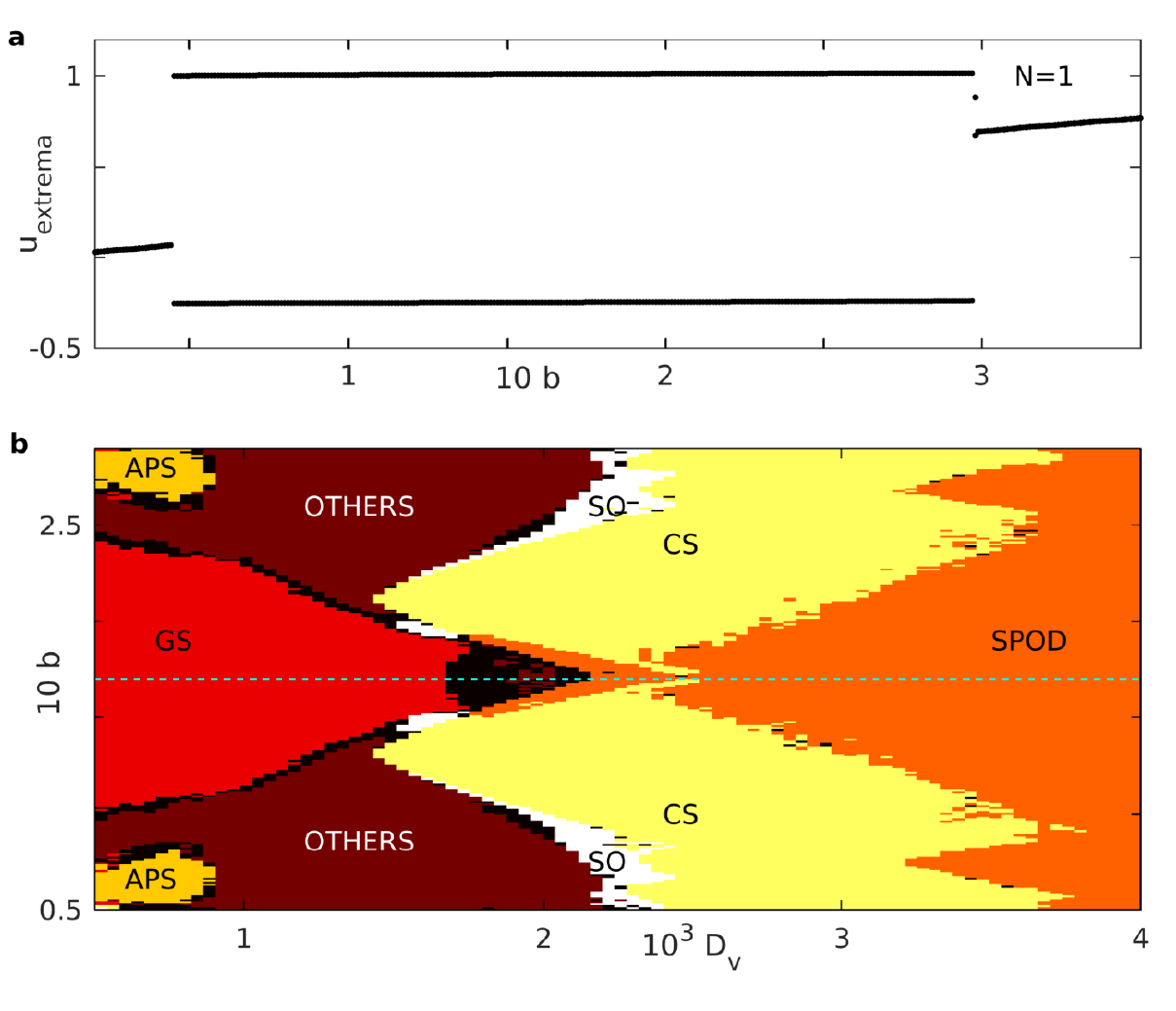}
\end{center}
\caption{
(a) Bifurcation diagram showing the range of values of the parameter 
$b$ for which a single FHN unit ($N=1$) exhibits oscillatory 
behavior. (b) Different dynamical regimes of a ring of $N=20$ FHN 
oscillators, diffusively coupled through the inactivation variable $v$ 
for the case $D_{u}=0$, are shown in the $D_{v}-b$ parameter plane 
over the range of values of $b$ for which a single unit oscillates. 
The dynamical regimes are labelled by the attractor to which the 
majority ($> 50\%$) of initial conditions converge, viz. SO, GS, APS, 
CS, SPOD and OTHERS. The dashed blue line divides the parameter space 
diagram into two halves that are found to be mirror images of each 
other. Note that the lower half of the parameter space diagram is the 
same as that shown in Fig.~2(b) of the main text. 
}
\label{SI2}
\end{figure}

For the bulk of our simulations, we have considered perhaps the most 
generic model of relaxation oscillators, viz., the Fitzhugh-Nagumo 
(FHN) model. Here, each oscillator is described by a fast activation 
variable $u$ and a slow inactivation variable $v$. As discussed in the 
main text, the time-evolution of this model is governed by the 
functions:
\begin{equation}
\begin{split}
\mathcal{F}(u, v) &= u(1 - u)(u - \alpha) - v\,,\\
\mathcal{G}(u, v) &= \epsilon (k u - v - b)\,,
\end{split}
\end{equation}
where $\alpha = 0.139$ and $k = 0.6$ are parameters specifying the 
kinetics, $b$ characterizes the asymmetry (related to the ratio of the 
time spent in the high- and low-value branches of the 
$u$ nullcline) and $\epsilon = 10^{-3}$ is the recovery rate.
Fig.~\ref{SI2}~(a) shows that in the absence of diffusive coupling
with its neighbors, oscillation is seen over a range of values of
the parameter $b$. Remaining within this interval, we have varied both
$b$ and the coupling strength $D_v$ to observe a range of collective
dynamics [Fig.~\ref{SI2}~(b)].

\clearpage
\section{Coarsening}
\begin{figure}[h!]
\centering
\includegraphics[width=0.95\columnwidth]{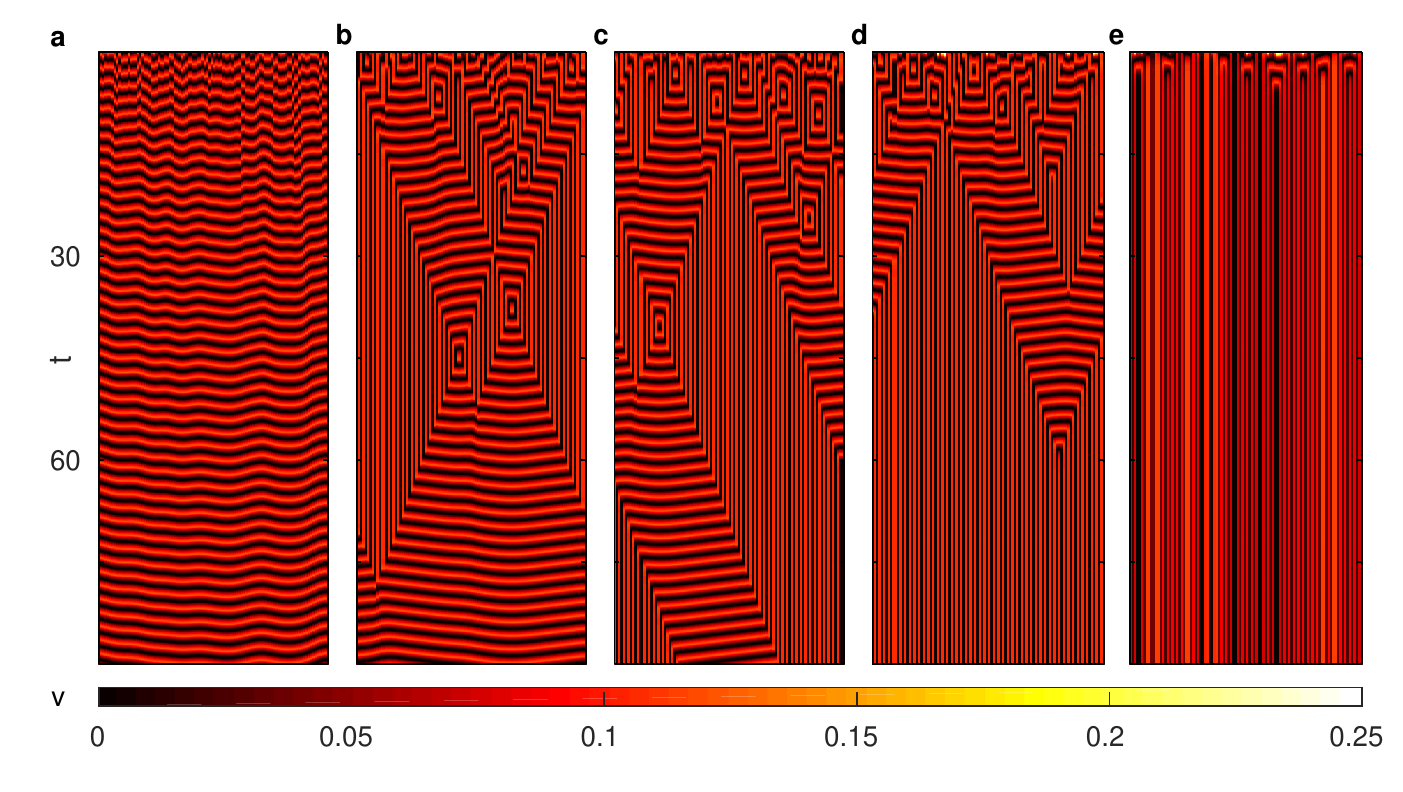}
\caption{
Pseudocolor representation of the spatiotemporal evolution of the 
activation variable $u$ on a ring of N=100 FHN oscillators. For a 
fixed value of the system parameter $b(= 0.175)$) and starting from 
random initial states, the system converges to different attractors as 
the value of $D_v$ is varied. In (a) and (e) the system quickly 
converges to GS ($D_v=7\times 10^{-4}$) and SPOD ($D_v=4\times 
10^{-3}$) respectively. (b-d) For an intermediate value of $D_{v} (=2\times
10^{-3})$, we
observe that the system can converge to GS (b),
exhibit an extremely long-lived state with coexisting GS and SPOD regions (c) or 
attain SPOD state (d) through a coarsening process where initially small 
domains having distinct dynamics 
merge with neighboring domains over time.
}
\label{SI3}
\end{figure}

In the main text we have shown that competition between attractors
at the boundary of the GS and SPOD regimes for a ring of FHN oscillators
results in coarsening-like
dynamics. Beginning with many small domains, each consisting of exclusively
oscillating or non-oscillating
elements, the collective dynamics exhibits a reduction in the
number of domains over time through a process of coalescence. Here we show
how the coarsening depends on the location of the system in the $D_v-b$
parameter space
by gradually increasing the coupling strength $D_v$ (keeping the
value of $b$ fixed at $0.175$). In Fig.~\ref{SI3}, starting from deep 
within the GS regime
[panel (a)] and terminating deep in the SPOD regime [panel(e)], 
we show spatiotemporal evolution of the inactivation 
variable $v$. We see that, for lower values of diffusion 
strength $D_v$, gradient synchronization (GS) occurs almost
immediately with negligible transients and hence, no 
coarsening is seen for this regime [panel (a)]. 
However, on increasing $D_v$ coarsening phenomena is observed in the
transient period leading to the asymptotic state (either GS or SPOD).
Indeed, we note that
coarsening is seen over a range of $D_v$, with a lower bound is 
$\sim 1.76 \times 10^{-3}$ and upper bound is $\sim 2 \times 10^{-3}$.
Panels (b-d) show coarsening to GS, extremely long-lived transients and 
SPOD for the same value of $D_v$ that falls in the above range.
Panel (e) shows that for high $D_v$ convergence to SPOD occurs rapidly
(without an appreciable period of transients) for high $D_v$.
 
\clearpage
\section{Robustness of Collective Dynamics: Results for $\bf {D_u = D_v}$}

\begin{figure}[h!]
\centering
\includegraphics[width=0.95\columnwidth]{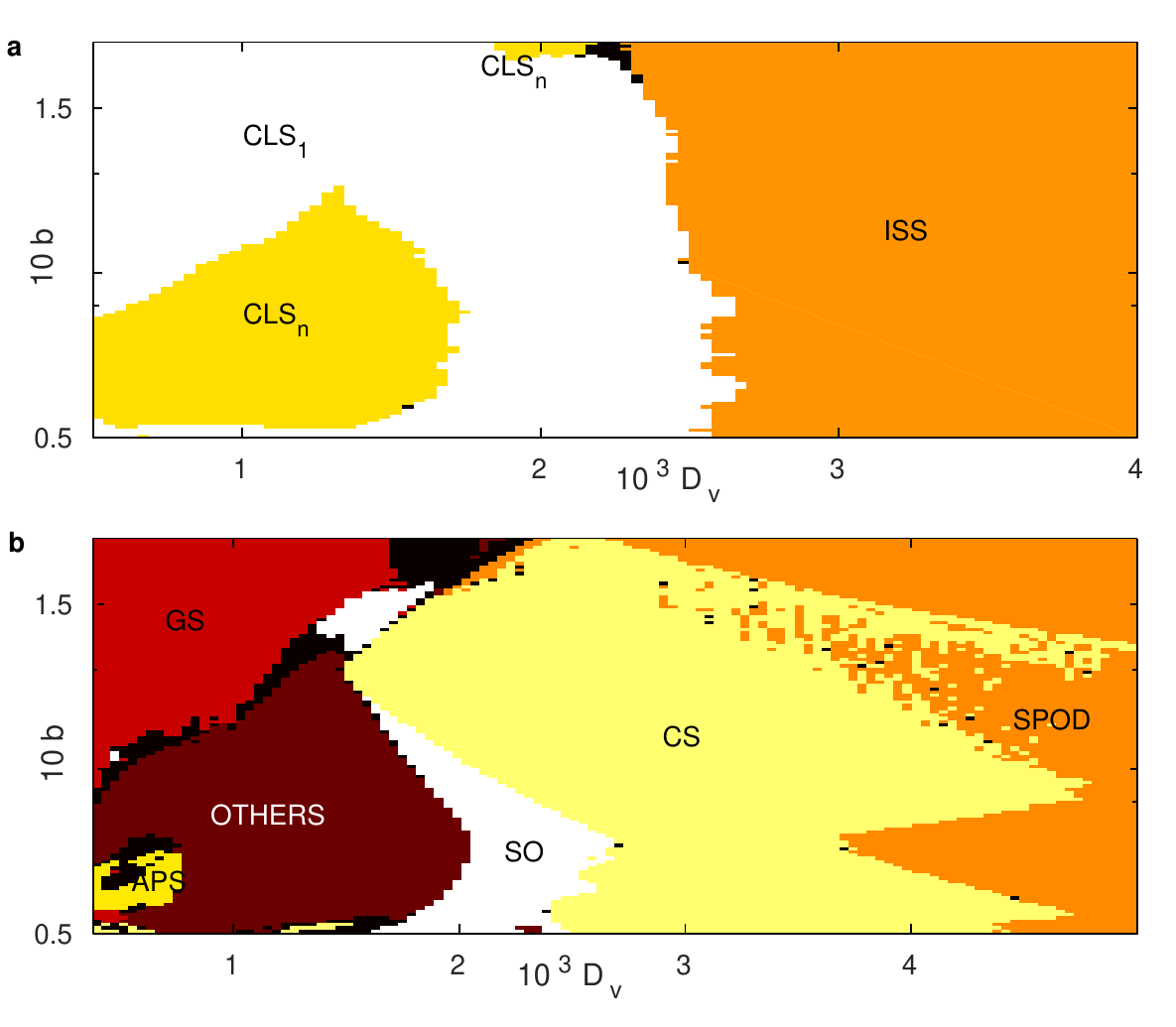}
\caption{
Different dynamical regimes of a system of FHN oscillators, 
diffusively coupled through the inactivation variable $v$ for the case 
$D_{u}=D_{v}$, are shown in the $D_{v}-b$ parameter plane for (a) a 
globally coupled system ($N=100$) that exhibits two fundamental 
patterns of collective activity, viz. Cluster Synchronization (CLS) 
and Inhomogeneous Steady State (ISS), and (b) a ring of size $N=20$ 
that exhibits SO, GS, APS, CS, SPOD and OTHERS. In each case, the 
dynamical regimes are labelled by the attractor to which the majority 
($> 50\%$) of initial conditions converge. 
}
\label{SI4}
\end{figure}

For the simulations reported in the main text, we have considered $D_u = 0$. 
However, as shown in Fig.~\ref{SI4}, we obtain qualitatively similar results 
even for the case $D_u = D_v$, i.e., when both activation and inactivation
variables are diffusively coupled with the same strength to their neighboring
sites. 

\end{document}